\documentclass{llncs}

\usepackage{graphicx}
\usepackage{algorithm}
\usepackage{algpseudocode}
\usepackage{url}
\usepackage{amssymb}
\usepackage{amsmath}

\begin{document}

\pagestyle{headings}
\title{MORE: Merged Opinions Reputation Model}

\author{Nardine Osman\inst{1} \and Alessandro Provetti\inst{2}
Valerio Riggi\inst{2} \and Carles Sierra\inst{1}}
\institute{Artificial Intelligence Research Institute (IIIA-CSIC), Barcelona, Spain
\and
Department of Math \& Informatics, University of Messina, Italy}

\maketitle              

\begin{abstract}
Reputation is generally defined as the opinion of a group on an aspect of a thing.  This paper presents a reputation model that follows a probabilistic modelling of opinions based on three main concepts: (1) the value of an opinion decays with time, (2) the reputation of the opinion source impacts the reliability of the opinion, and (3) the certainty of the opinion impacts its weight with respect to other opinions.  Furthermore, the model is flexible with its opinion sources: it may use explicit opinions or implicit opinions that can be extracted from agent behaviour in domains where explicit opinions are sparse.  We illustrate the latter with an approach to extract opinions from behavioural information in the sports domain, focusing on football in particular.  One of the uses of a reputation model is predicting behaviour.  We take up the challenge of predicting the behaviour of football teams in football matches, which we argue is a very interesting yet difficult approach for evaluating the model.  
\keywords{Trust, reliability and reputation}
\end{abstract}

\section{Introduction}
This paper is concerned with the classic, yet crucial, issue of reputation. We propose MORE, the \textsl{Merged Opinions REputation model}, to compute reputation on the basis of opinions collected over time.   
MORE uses a probabilistic modelling of reputation; adopts the notion of information decay; considers the reliability of an opinion as a function of the reputation of the opinion holder; and assesses the weight of an opinion based on its certainty.  This latter feature constitutes the most novel feature of our algorithm.  

Furthermore, MORE may be applied to fields with varying abundancy of explicit opinions available.  
In other words, if explicit opinions are available, as it is the case with so-called {\em eMarkets,} then those opinions may directly be used by MORE. 
In other cases, where such opinions are sparse, behavioural information can be translated into opinions that MORE can then use.   
For example, if Barcelona beats Real Madrid at football, then this may be translated into {\em mutual opinions} where Barcelona expresses Real Madrid's inadequate skills and Real Madrid acknowledges Barcelona's superior skills.  
This paper also proposes an approach for extracting opinions from behavioural information in the sports domain.  

MORE's calculated reputation measures may then be used for different objectives, from ranking performance to predicting behaviour and sports results. 

Evaluating reputation is a notoriously tricky task, since there seldom is an objective measure to compare to.  
For instance, how can we prove which opinion is correct and which is biased?  
In this paper, we present an extensive validation effort that has sought to assess MORE's predictive abilities in the football domain, where accurate predictions are notoriously hard to make~\cite{hill1974association}.



The rest of this paper is divided as follows: Section~\ref{sec:more} presents the MORE model, 
Section~\ref{sec:more_approximation} introduces the necessary approximations, 
Section~\ref{sec:more_algorithm} summarises the MORE algorithm; 
Sections~\ref{sec:scores2ops}, \ref{sec:predictions}, and~\ref{sec:soccer+chess} presents our evaluation, before concluding with Section~\ref{sec:conclusion}.
\section{The MORE Model}\label{sec:more}
We define the opinion that agent $\beta$ may form about agent $\alpha$ at time $t$ as: $o_{\beta}^{t}(\alpha)=\{e_{1}\!\mapsto\!v_{1},\dots,e_{n}\!\mapsto\!v_{n}\}$, where $G=\{\alpha,\beta,\dots\}$ is a set of agents; $t\in T$ and $T$ represents calendar time; $E=\{e_1,\dots,e_n\}$ is an ordered evaluation space where the terms $e_i$ may account for terms such as \emph{bad}, \emph{good}, \emph{very good} and so on; and $v_{i}\in [0,1]$ represents the value assigned to each element $e_{i}\in E$ under the condition that $\sum_{i\in[1,|E|]} v_{i}=1$. 
In other words, the opinion $o_{\beta}^{t}(\alpha)$ is specified as a discrete probability distribution over the evaluation space $E$. 
We note that the opinion one holds with respect to another may change with time, hence various instances of $o_{\beta}^{t}(\alpha)$ may exist for the same agents $\alpha$ and $\beta$ but for distinct time instants $t$.

Now assume that at time $t$, agent $\beta$ forms an opinion $o_{\beta}^{t}(\alpha)$ about agent $\alpha$. 
To be able to properly interpret the opinion, we need to consider how reliable $\beta$ is in giving opinions.  
We reckon that the overall reliability of any opinion is the reliability of the person holding this opinion, which changes along time.  That is the more reliable an opinion is, the closer its reviewed value is to the original one; inversely, the less reliable an opinion is, the closer its reviewed value is to the flat (or uniform) probability distribution $\mathbb{F}$, which represents complete ignorance and is defined as $\forall \, e_{i}\in E \;\cdot\;\mathbb{F}(e_{i})=1/|E|$.  This reliability value $\mathcal{R}$ is defined later on in Section~\ref{sec:reputation}. 
However, in this section, we use this value to assess the reviewed value $\mathbb{O}_{\beta}^{t}(\alpha)$ of the expressed opinion $o_{\beta}^{t}(\alpha)$, which we define accordingly:
\begin{equation}\label{eq:opinion}
\mathbb{O}_{\beta}^{t}(\alpha) = {\cal R}^{t}_\beta\times o_\beta^{t}(\alpha) + (1-{\cal R}^{t}_\beta)\times\mathbb{F}
\end{equation}

\subsection{Opinion Decay}
Information loses its value with time. 
Opinions are no exception, and their integrity decreases with time as well. 
Based on the work of~\cite{sierra_icore09}, we say the value of an opinion should tend to ignorance, which may be represented by the flat distribution $\mathbb{F}$. %
In other words, given a distribution $\mathbb{O}^{t'}$ created at time $t'$, we say at
time $t>t'$, $\mathbb{O}^{t'}$ would have decayed to
$\mathbb{O}^{t}=\Lambda(t,\mathbb{F},\mathbb{O}^{t'})$, where
$\Lambda$ is the \emph{decay function} satisfying the property
$\lim_{t'\to\infty}\mathbb{O}^{t'}=\mathbb{F}$. 

One possible definition, used by MORE, for $\Lambda$ is the following:

\begin{equation}\label{eq:decay}
  \mathbb{O}^{t'\rightarrow t}= \nu^{\Delta_{t}}\; \mathbb{O}^{t'}+ (1-\nu^{\Delta_{t}}) \;\mathbb{F}
\end{equation}

\noindent
where $\nu \in [0,1]$ is the decay rate, and:

\begin{equation}
{\Delta_{t}} =
\begin{cases}
  0 \textrm{ , if $t-t'< \kappa$}\\
  1 + \frac{t - t'}{\kappa} \textrm{ , otherwise}
\end{cases}
\end{equation}

$\Delta_{t}$ serves the purpose of establishing a minimum {\it grace} period during which the information does not decay and that once reached the information starts decaying. 
This period of grace is determined by the parameter $\kappa$, which is also used to control the pace of decay.

\subsection{Certainty and its Impact on Group Opinion}
A group opinion on something at some moment is based on the aggregation of all the previously-expressed individual opinions. 
However, the certainty of each of these individual opinions has a crucial impact on the aggregation.  This is a concept that, to our knowledge, has not been used in existing aggregation methods for reputation.  
We say, the more uncertain an opinion is then the smaller its effect on the final group opinion is. 
The maximum uncertainty is defined in terms of the flat distribution $\mathbb{F}$.  Hence, we define this certainty measure, which we refer to as the opinion's value of
information, as follows:

\begin{equation}\label{eq:opinion_info}
{\cal I}(\mathbb{O}_{\beta}^{t}(\alpha)) = {\cal
  H}(\mathbb{O}_{\beta}^{t}(\alpha)) - {\cal H}(\mathbb{F})
\end{equation}

where, ${\cal H}$ represents the entropy of a probability distribution, or the value of information of a probability distribution. 
In other words, the certainty of an opinion is the difference in entropies of the opinion and the flat distribution.

Then, when computing the group opinion, we say that any agent can give opinions about another at different moments in time. 
We define $T_\beta(\alpha) \subseteq T$ to describe the set of time points at which $\beta$ has given opinions about $\alpha$. 
The group opinion about $\alpha$ at time $t$, $\mathbb{O}_{G}^{t}(\alpha)$, is then calculated as follows:

\begin{equation}\label{eq:group_opinion}
\mathbb{O}_{G}^{t}(\alpha) = \frac{\displaystyle\sum_{\beta \in G} \sum_{t' \in
    T_\beta(\alpha)}\mathbb{O}_{\beta}^{t' \to t}(\alpha) \cdot {\cal
    I}(\mathbb{O}_{\beta}^{t' \to t}(\alpha)) }{\displaystyle\sum_{\beta \in
    G}\sum_{t' \in T_\beta(\alpha)} {\cal I}(\mathbb{O}_{\beta}^{t' \to
    t}(\alpha))}
\end{equation}

This equation states that the group opinion is an aggregation of all the decayed individual opinions $\mathbb{O}_{\beta}^{t'\to t}(\alpha)$ that represent the view of
every agent $\beta$ that has expressed an opinion about $\alpha$ at
some point $t'$ in the past.  
However, different views are given different weights, depending on the value of their information $\mathcal{I}(\mathbb{O}_{\beta}^{t' \to t}(\alpha))$.

Note that in the proposed approach, one's latest opinion does not override previous opinions.  
This choice to override previous opinions or not is definitely context dependent.  For example, consider one providing an opinion about a certain product on the market, then changing his opinion after using the product for some time. 
In such a case, only the latest opinion should be considered and it should override the previous opinion. 
However, in our experiments, we use the sports domain, where winning football matches are interpreted as opinions formed by the teams about each others strength in football. 
In such a case, the opinions obtained from the latest match's score should not override opinions obtained from previous matches. 
In such a context, past opinions resulting from previous matches will still need to be considered when assessing a team's reputation.

Finally, we note that initially, at time $t_{0}$, we have $\forall \, \alpha \in
G \;\cdot\; \mathbb{O}_{G}^{t_{0}}(\alpha)=\mathbb{F}$. 
In other words, in the absence of any information, the group opinion is equivalent to the flat distribution accounting for maximum ignorance. 
As individual opinions are expressed, the group opinion starts changing following Equation~\ref{eq:group_opinion}.  

\subsection{Reliability and Reputation}\label{sec:reputation}
An essential point in evaluating the opinions held by someone is considering how reliable they are. 
This is used in the interpretation of the opinions issued by agents (Equation~\ref{eq:opinion}). 
The idea behind the notion of reliability is very simple. 
A person who is considered very good at solving a certain task, i.e. has a high reputation with respect to that task, is usually considered an expert in \emph{assessing} issues related to that task. 
This is a kind of {\it ex-cathedra} argument. 
An example of current practice supported by this argument is the selection of members of committees or advisory boards.

But how is reputation calculated?  
First, given an evaluation space $E$, it is easy to see what could be the best opinion about someone: the `ideal' distribution, or the `target', which is defined as $\mathbb{T} = \{e_n\mapsto 1\}$, where $e_{n}$ is the top term in the evaluation space.  
Then, the reputation of $\beta$ within a group $G$ at time $t$ may be defined as the distance between the  current aggregated opinion of the group $\mathbb{O}_G^{t}(\beta)$ and the ideal distribution $\mathbb{T}$, as follows: 

\begin{equation}\label{eq:reputation}
  {\cal R}^{t}_\beta = 1 - emd(\mathbb{O}_G^{t}(\beta), \mathbb{T})
\end{equation}

where $emd$ is the earth movers distance that measures the distance between two probability distributions~\cite{emd} (although 
other distance measurements may also be used). %
The range of the \textit{emd} function is [0,1], where $0$ represents the minimum distance (i.e. both distributions are identical) and $1$ represents the maximum distance possible between the two distributions.

As time passes and opinions are formed, the reputation measure evolves along with the group opinion. 
Furthermore, at any moment in time, the measure $\mathcal{R}^{t}$ can be used to rank the different agents as well as assess their reliability.
\section{Necessary Approximation}\label{sec:more_approximation}
As Equation~\ref{eq:group_opinion} illustrates, the group opinion is
calculated by aggregating the decayed individual opinions and
normalising the final aggregated distribution by considering the value
of the information of each decayed opinion
($\mathcal{I}(\mathbb{O}_{\beta}^{t'\to t}(\alpha))$).  
This approach imposes severe efficiency constraints as it demands exceptional computing power: each time the group opinion needs to be calculated, all past opinions need to decay to the time of the request, and the value of the information of these decayed opinions should be recomputed.  

We suggest an approximation to Equation~\ref{eq:group_opinion} that allows us to apply the
algorithm over a much longer history of opinions.  
To achieve this, when a group opinion is requested, its value is calculated by obtaining the latest group opinion and decaying it accordingly. 
In other words, we assume the group opinion to decay just like any other
source of information.  
Instead of recalculating them over and over again, we simply decay the latest calculated value following Equation~\ref{eq:decay} as follows:

$$\mathbb{O}_{G}^{t}(\alpha)= \nu^{\Delta_{t}}\; \mathbb{O}_{G}^{t'}(\alpha)+ (1-\nu^{\Delta_{t}}) \;\mathbb{F}$$

When a new opinion is added, the new group opinion is then updated by adding the new opinion to the decayed group opinion. 
In this case, normalisation is still achieved by considering the value of the information of the opinions being aggregated; however, it also
considers the number of opinions used to calculate the latest group
opinion. 
This is because one new opinion should not have the exact weight as all the previous opinions combined.  
In other words, more weight should be given to the group opinion, and this weight should be based on the number of individual opinions contributing to that group opinion.
As such, when a new opinion $o_{\beta}^{t}(\alpha)$ is added, Equation~\ref{eq:group_opinion} is replaced with Equation~\ref{eq:group_opinion_app}:

\begin{equation}\label{eq:group_opinion_app}
  \mathbb{O}_{G}^{t}(\alpha) =
  \frac{n_{\alpha}\,\mathbb{O}_{G}^{t'\to t}(\alpha) \cdot {\cal I}(\mathbb{O}_{G}^{t'\to t}(\alpha))
    + \mathbb{O}_{\beta}^{t}(\alpha) \cdot {\cal I}(\mathbb{O}_{\beta}^{t}(\alpha))}{
    n_{\alpha}\,{\cal I}(\mathbb{O}_{G}^{t'\to t}(\alpha)) + {\cal I}(\mathbb{O}_{\beta}^{t}(\alpha))}
\end{equation}

where $n_{\alpha}$ represents the number of opinions used to calculate the group opinion about $\alpha$.

Of course, this approach provides an approximation that is not equivalent to the exact group opinion calculated following
Equation~\ref{eq:group_opinion}.  
This is mainly because the chosen decay function (Equation~\ref{eq:decay}) is not a linear function since the decay parameter $\nu$ is raised to the exponent of $\Delta_{t}$, which is time dependent.  
In other words, decaying the group opinion as a whole results in a different probability distribution than decaying all the individual opinions separately and aggregating the results following Equation~\ref{eq:group_opinion}.
Hence, there is a need to know how close is the approximate group opinion to the exact one.  
In what follows, we introduce the test used for comparing the two, along with the results of this test.

\subsection{The Approximation Test}
To test the proposed approximation, we generate a number of random opinions 
$\mathbb{O}_{\beta_{i}}^{t}(\alpha)$ over a number of years, where $\alpha$ is fixed,
$\beta_{i}$ is an irrelevant variable (although we do count the number
of opinion sources every year, the identity of the source itself is
irrelevant in this specific experiment), and $t$ varies according to the
constraints set by each experiment.  For example, if 4 opinions were
generated every year for a period of 15 years, then the following is
the set of opinion sets that will be generated over the years:

$$\{\{\mathbb{O}_{\beta_{1}}^{1}(\alpha), ...,
\mathbb{O}_{\beta_{4}}^{1}(\alpha)\}, ...,
\{\mathbb{O}_{\beta_{1}}^{15}(\alpha), ...,
\mathbb{O}_{\beta_{4}}^{15}(\alpha)\}\}$$

With every generated opinion, the group opinion is calculated
following both the exact model (Equation~\ref{eq:group_opinion}) and
the approximate model (Equation~\ref{eq:group_opinion_app}).  We then
plot the distance between the exact group opinion and the
approximate one.  
The distance between those two distributions is
calculated using the earth mover's distance method outlined earlier.  We note that a good approximation is an approximation where the earth mover's distance (EMD) is close to 0.

Two different experiments were executed.
In the first, 10 opinions were being generated every year over a
period of 6 years.  In the second, 4 opinions where being generated
every year over a period of 15 years.  Each of these experiments were  
repeated several times to test a variety of decay parameters.  The
final results of these experiments are presented in the following
section.

\subsection{Results of the Approximation Test}
\begin{figure}[!ht]
  \centering
  \includegraphics[scale=0.26]{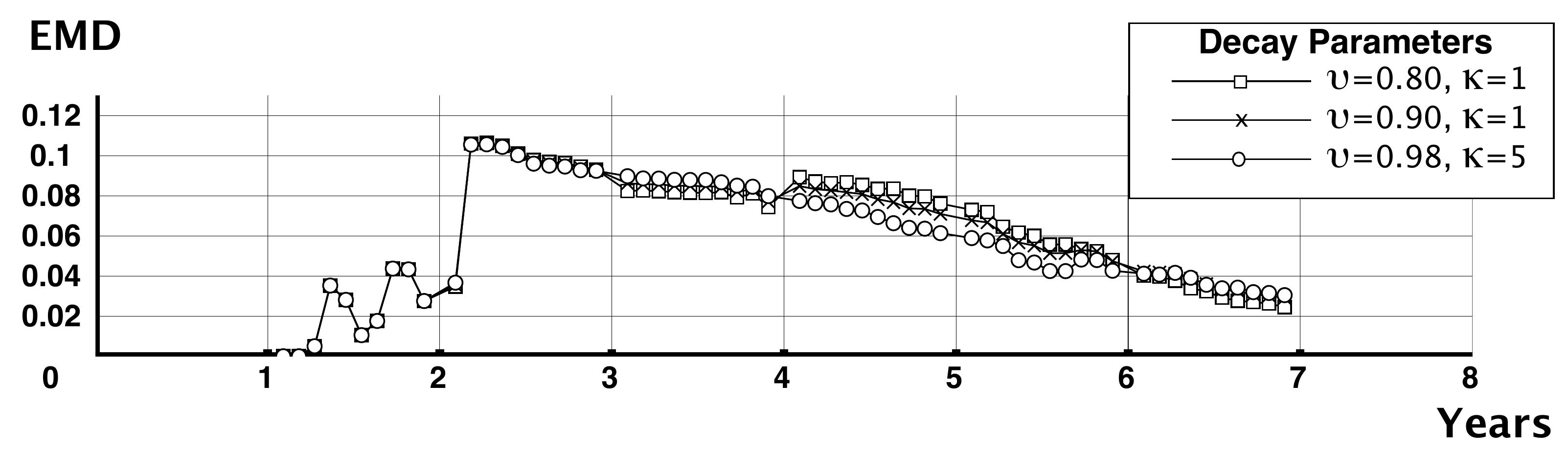}
  \caption{Distance between the exact and approximate $\mathbb{O}_{G}^{t}$}
  \label{fig:approximation_results}
\end{figure}

Figure~\ref{fig:approximation_results} presents the results of the
first experiment introduced above.  The results show that the
approximation error increases to around $11\%$ in the first few
rounds, and after 12 opinions have been introduced.  The approximation
error then starts to decrease steadily until it reaches $0.3\%$ when
60 opinions have been added.  Experiment 2 has the exact same results,
although spanning over 15 years instead of 6.  For this reason, as
well as well as lack of space, we do not present the second
experiment's results here.  However, we point that both experiments
illustrate that it is the number of opinions that affect 
the increase/decrease in the EMD distance, rather than the number of
years and the decay parameters.  In fact, undocumented results
illustrate that the results of Figure~\ref{fig:approximation_results}
provide a good estimate of the worst case scenarios, since the earth
mover's distance does not grow much larger for smaller $\nu$ values,
but starts decreasing towards $0$.  When $\nu=0$ and the decay is
maximal (i.e. opinions decay to the flat distribution at every
timestep), the EMD distance is $0$.  However, when the decay is
minimal (i.e. opinions never decay), then the results are very close
to the case of $\nu=0.98$ and $\kappa=5$.

  

We conclude that the larger the available number of opinions, then the more precise the approximation is.  
This makes this approximation suitable for applications where more and more opinions are available.

\section{The MORE Algorithm}\label{sec:more_algorithm}
The merged opinions reputation model, MORE, is implemented using
the approximation of Section~\ref{sec:more_approximation} and formalised by Algorithm~\ref{alg:more}.

\begin{algorithm}
  \caption{The MORE Algorithm}
  \label{alg:more}
  \small
  \begin{algorithmic}
    \Require $E = \{e_1, \dots, e_n\}$ to be an evaluation space
    \Require $G = \{\alpha, \beta, \dots\}$ to be a group of agents
    \Require $t \in \mathbb{N}$ to be a point in time
    \Require {\sc odb} to describe the database of all opinions
    \Require $o_{\beta}^{t'}(\alpha) \preceq o_{\delta}^{t}(\gamma) = \{\top \hbox{, if $t' \preceq t$}; \bot \hbox{, otherwise}\}$
    \Require $\mathbb{O}_{X}^{t' \to t}(\alpha) = (\mathbb{O}_{X}^{t'}(\alpha) - \mathbb{F})\nu^{1+(t-t')/\kappa} + \mathbb{F}$, where $\nu\in[0,1]$ is the decay parameter and $\kappa \in \mathbb{N}$ is the pace of decay
    \Require $emd: 2^{\mathbb{P}(E)} \times 2^{\mathbb{P}(E)} \to [0,1]$ to represent the earth mover's distance function that calculates the distance between two probability distributions

    \State $\forall \, e_i \in E \;\cdot\; \mathbb{F}(e_i) = 1/n$
    \State $\forall \, e_i \in E \;\cdot\; (i<|E| \Rightarrow \mathbb{T}(e_i) = 0) \wedge (i=|E| \Rightarrow \mathbb{T}(e_i) = 1)$
    
    \State ${\cal H}(\mathbb{F}) =  - \log (1/n)$
    
    \State $\forall \,\alpha \in G \;\cdot\; \mathbb{O}_{G}^{t_{0}}(\alpha)=\mathbb{F}$
    \State $\forall \,\alpha \in G \;\cdot\; n_{\alpha} = 0$
    
    \While{$\exists \, o_{\beta}^{t}(\alpha) \in \hbox{{\sc odb}} \;\cdot\;
      (\forall \, o \in \hbox{{\sc odb}} \;\cdot\; o_{\beta}^{t}(\alpha) \preceq o)$}

    \State $ {\cal R}^{t}_\beta = 1 - emd(\mathbb{O}_G^{t'\to t}(\beta), \mathbb{T})$
    \State $\mathbb{O}_{\beta}^{t}(\alpha) = {\cal R}^{t}_\beta\times o_\beta^{t}(\alpha) + (1-{\cal R}^{t}_\beta)\times\mathbb{F}$

    \State ${\cal I}(\mathbb{O}_{\beta}^{t}(\alpha)) = - \displaystyle\sum_{e_i \in E} \mathbb{O}_{\beta}^{t}(\alpha)(e_i) \cdot \log \mathbb{O}_{\beta}^{t}(\alpha)(e_i) - {\cal H}(\mathbb{F})$
    \State ${\cal I}(\mathbb{O}_{G}^{t' \to t}(\alpha)) \!=\! - \!\!\!\displaystyle\sum_{e_i \in E} \!\!\mathbb{O}_{G}^{t'\to t}(\alpha)(e_i) \cdot \log \mathbb{O}_{G}^{t'\to t}(\alpha)(e_i) - {\cal H}(\mathbb{F})$

    \State $\mathbb{O}_{G}^{t}(\alpha) = \frac{n_{\alpha}\,\mathbb{O}_{G}^{t'\to t}(\alpha)\cdot {\cal I}(\mathbb{O}_{G}^{t'\to t}(\alpha)) \;+\; \mathbb{O}_{\beta}^{t}(\alpha)\cdot {\cal I}(\mathbb{O}_{\beta}^{t}(\alpha))}{n_{\alpha}\,{\cal I}(\mathbb{O}_{G}^{t'\to t}(\alpha)) \;+\; {\cal I}(\mathbb{O}_{\beta}^{t}(\alpha))}$

    \State $ {\cal R}^{t}_\alpha = 1 - emd(\mathbb{O}_G^{t}(\alpha),
    \mathbb{T})$

    \State $n_{\alpha}=n_{\alpha}+1$
    \EndWhile
  \end{algorithmic}
\end{algorithm}

In summary, the algorithm is called with a predefined set of opinions,
or the opinions database {\sc odb}.  For each opinion in {\sc odb}, the
reviewed value is calculated following Equation~\ref{eq:opinion}, the
informational value of the opinion as well as that of the decayed
latest group opinion are calculated following
Equation~\ref{eq:opinion_info}, the updated group opinion is then
calculated following Equation~\ref{eq:group_opinion_app}, and the
reputation of the agent is calculated via
Equation~\ref{eq:reputation}.  These steps are repeated for all
opinions in {\sc odb} in an ascending order of time, starting from the
earliest given opinion and moving towards the latest given opinion.  

We note that the complexity of this algorithm is constant ($\mathcal{O}(1)$).  Whereas if we were using Equation~\ref{eq:group_opinion} as opposed to the proposed approximation, then the complexity would have been linear w.r.t. the number of opinions $n$ ($\mathcal{O}(n)$).  For very large datasets, such as those used in the experiment of Section~\ref{sec:soccer+chess}, the approximation does provide a great advantage.
\section{From Raw Scores to Opinions}\label{sec:scores2ops}
This section describes the extraction of opinions from behavioural information.  
While we focus on football, we note that these methods may easily be applied to other domains. 
%
We say the possible outcomes of a match between teams $\alpha$ and $\beta$ are as follows: {\em (i)} $\alpha$ wins, {\em (ii)} $\alpha$ loses, or {\em (iii)} the match ends up in a draw.
We denote as $ng(\alpha)$ (resp., $ng(\beta)$) the number of goals scored by $\alpha$ (resp., $\beta$).
We then define three methods to convert match results into opinions.
Generated opinions belongs to a binary evaluation space consisting of two outcomes, namely {\tt bad} ($B$) and {\tt good} ($G$): $E=\{B,G\}$.

\subsection{The Naive Conversion}\label{sub:naive}
In this first strategy, we simply look for the winner.
If $\alpha$ wins, then it receives an opinion from $\beta$ equal to $o_{\beta}^{t}(\alpha)=\{B\!\mapsto\! 0, G\!\mapsto\!1\}$, and $\beta$ will get an opinion from $\alpha$ equal to $o_{\alpha}^{t}(\beta)=\{B\!\mapsto\! 1 ,G\!\mapsto\!0\}$.
In case of a draw, they both get the same opinion: $o_{\beta}^{t}(\alpha) = o_{\alpha}^{t}(\beta) = \{B\!\mapsto\! 0.5,G\!\mapsto\! 0.5\}$.
The method is quite simple and it does not take into account important aspects such as the final score of the match. 
For instance, losing 0 to 3 is equivalent to losing 2 to 3.

\subsection{Margin-of-Victory Conversion}\label{sub:margin}
A second strategy we consider is called {\em Margin of Victory~--~MV}.
The margin of victory of a match involving clubs $\alpha$ and $\beta$ is defined as the difference of goals $M = ng(\alpha) - ng(\beta)$ scored by $\alpha$ and $\beta$. Of course $M >0$ if $\alpha$ wins.
The main idea here is this: if we know $\alpha$ beats $\beta$, this tells us something about the relative strength of $\alpha$ against $\beta$.
If we know $\alpha$ scored more than 3 goals against $\beta$ (which is rather unusual in many professional leagues), we could probably have a better picture of the relative strength of the two clubs.
We believe that including more data in the process of generating opinions should produce more accurate results and, ultimately, this should help us in better predicting the outcome of a football match.
The rules we used to include the number of goals scored by each club are as follows:

\begin{equation}\label{eq:scoreop2}
o_{\alpha}^{t}(\beta)= 
\begin{cases}
\left\{B\!\mapsto\! 0.5 ,G\!\mapsto\! 0.5\right\} & \text{, for a 0-0 tie}\\
\left\{B\!\mapsto\! \frac{ng(\alpha)}{ng(\alpha) + ng(\beta)}, G\!\mapsto\! \frac{ng(\beta)}{ng(\alpha) + ng(\beta)}  \right\} & \text{, otherwise}
\end{cases}
\end{equation}

In analogous fashion we can compute the opinion of $\beta$ on $\alpha$.
Equation \ref{eq:scoreop2} tells us that if the margin of victory  $ng(\alpha) - ng(\beta)$ is large, then $ng(\alpha)$ is higher than  $ng(\beta)$ and the ratio $\frac{ng(\alpha)}{ng(\alpha) + ng(\beta)}$ will be closer to 1.
As a consequence, the larger the margin of victory between $\alpha$ and $\beta$, the more likely $\alpha$ will get an evaluation biased towards {\tt good.}
In case of a 0-0 tie, the terms $\frac{ng(\alpha)}{ng(\alpha) + ng(\beta)}$ and $\frac{ng(\beta)}{ng(\alpha) + ng(\beta)}$ are undefined.
To manage such a configuration, we assume that the probability that $\alpha$ (resp., $\beta$) gets the evaluation {\tt good} is equal to the probability it gets the evaluation {\tt bad.}

A potential drawback of the {\em MV} strategy is that different scores may be translated into the same distribution.  This happens every time one of the clubs does not score any goal.  For instance, the winners in two matches that end with the scores $1-0$ and $4-0$ would received an opinion $\{B\!\mapsto\! 0,  G\!\mapsto\! 1\}$, as calculated by the {\em MV} strategy.

\subsection{Gifted Margin of Victory}\label{sub:gifted-mv}
The third strategy we propose is called the {\em Gifted Margin of Victory} -- {\em GMV}.  
It has been designed to efficiently handle the case of football matches in which one of the clubs does not score any goal.
The {\em GMV} strategy computes opinions accordingly:

\begin{equation}\label{eq:scoreop21}
\begin{array}{l}
o_{\beta}^{t}(\alpha)=\left\{ B\!\mapsto\! \frac{ng(\alpha)+X}{ng(\alpha) + ng(\beta) + 2X} ,  G\!\mapsto\! \frac{ng(\beta) + X}{ng(\alpha) + ng(\beta) + 2X}  \right\}
\end{array}
\end{equation}
\begin{equation}\label{eq:scoreop3}
\begin{array}{l}
o_{\alpha}^{t}(\beta)=\left\{ B\!\mapsto\! \frac{ng(\beta) + X}{ng(\alpha) + ng(\beta) + 2X} ,  G\!\mapsto\! \frac{ng(\alpha) + X}{ng(\alpha) + ng(\beta) + 2X} \right\}
\end{array}
\end{equation}

In other words, we give as a gift both clubs with a bonus of $X > 0$ goals in order to manage all matches in which one (or possibly both) of the two clubs does not score any goal.
Here $X$ is any positive real number.
If $X \rightarrow 0$, then the {\em GMV} strategy would collapse to the {\em MV} strategy.
On the other hand, if $X$ is extremely large then the constant $X$ would dominate over both $ng(\alpha)$ and $ng(\beta)$ and the terms $\frac{ng(\beta) + X}{ng(\alpha) + ng(\beta) + 2X}$ and $\frac{ng(\alpha)+X}{ng(\alpha) + ng(\beta) + 2X}$ would converge to 0.5.
This result is potentially negative because the probability that any team is evaluated as {\tt good} is substantially equivalent to the probability that it is evaluated as {\tt bad} and, therefore, all the opinions would be intrinsically uncertain.
An experimental analysis was carried out to identify the value of $X$ guaranteeing the highest prediction accuracy.
Due to space limitations we omit the discussion on the experimental tuning of the $X$ parameter and we suffice with the results of our experiment that show that the best value found for $X$ was 1.

A further improvement of the {\em GMV} strategy comes from normalization. 
Normalization is motivated by the observation that, since $X >0$, term $ng(\alpha) + X$ (resp., $ng(\beta) + X$) is strictly less than $ng(\alpha) + ng(\beta) + 2X.$ 
Hence, $\alpha$ (resp., $\beta$) will never get an opinion where the probability of {\tt good} comes close to 1, even if it has scored much more goals than $\beta$ (resp., $\alpha$). 
At the same time, since $ng(\alpha)+X>0$, there is no chance that $\alpha$ will get an opinion where the probability of {\tt bad} is close to 0.  

Let $p_{\alpha, \beta}^{GMV}(G)$ be the probability of $\alpha$ being evaluated {\tt good} by $\beta$, according to the $GMV$ strategy.  We then normalize $p_{\alpha, \beta}^{GMV}(G)$ to the [0,1] range by considering, for a given set $\mathcal{S}$ of teams, the highest and lowest probabilities of being evaluated {\tt good} according to the calculations of the GMV strategy, which we denote as $M(\mathcal{S})$ and $m(\mathcal{S})$, respectively. 
We then define the normalized probability $\hat{p}_G(\alpha)$ of team $\alpha$ being evaluated {\tt good} by $\beta$ as follows:

\begin{equation}\label{eq:GMV-normalization}
\hat{p}_G(\alpha) =\frac{p_{\alpha, \beta}^{GMV}(G) - m(\mathcal{S})}{M(\mathcal{S}) - m(\mathcal{S})}
\end{equation}
And the probability of team $\alpha$ being evaluation {\tt bad} by $\beta$ becomes: $\hat{p}_B(\alpha) = 1 - \hat{p}_G(\alpha)$.



\section{From Reputation to Predictions}\label{sec:predictions}
This section illustrates how we can use MORE to predict the outcome of a football match.
We note that a football match may be depicted as an {\em ordered pair} $\langle \alpha, \beta \rangle$, where $\alpha$ and $\beta$ are opponent clubs.
We will follow this convention: we will let $\alpha$ be the `home club' whereas $\beta$ will be the `visiting club'. 
To compute the reputation of teams $\alpha$ and $\beta$, we define the {\em relative strength} of $\alpha$ w.r.t. $\beta$ at time $t$ as follows:

\begin{equation}
\label{eqn:relative-strength}
r_{\alpha,\beta}(t) = \frac{{\cal R}^{t}_\alpha}{{\cal R}^{t}_\alpha + {\cal R}^{t}_\beta}
\end{equation}

In what follows, and for simplification, we omit the reference to time $t$ and we use the simplified notation $r_{\alpha,\beta}$.
Notice that $0 \leq r_{\alpha,\beta} \leq 1$ and the higher (resp., lower) $r_{\alpha,\beta}$ is, the stronger (resp., weaker) the club $\alpha$ is at playing and winning a football match.
We shall adopt the following rules to predict the outcome of a match:\footnote{We note that we look for values that are approximately greater ($\gtrapprox$), approximately less than ($\lessapprox$), or approximately equal ($\approx$) to $\frac{1}{2}$.  In practice, this is achieved by defining three different intervals to describe this.}

\begin{enumerate}
\item If $r_{\alpha,\beta} \gtrapprox \frac{1}{2}$, then the winner will be $\alpha$.
\item If $r_{\alpha,\beta} \approx \frac{1}{2}$, then the match will end up in a draw.
\item If $r_{\alpha,\beta} \lessapprox \frac{1}{2}$, then the winner will be $\beta$.
\end{enumerate}


\section{Experimental Results}\label{sec:soccer+chess}
In this section, we test the effectiveness of our approach.
In detail, we designed our experiments to answer the following questions:
\begin{enumerate}
\item[$\mathbf{Q}_1$.] What is the accuracy of the MORE algorithm in correctly predicting the outcome of a football match?

\item[$\mathbf{Q}_2$.] 
Which score-to-opinion strategy is reliably the most accurate?

\item[$\mathbf{Q}_3$.] To what extent does information decay impact the accuracy of MORE?

\end{enumerate}

\subsection{Datasets and Experimental Procedure}
To answer questions $\mathbf{Q}_1$--$\mathbf{Q}_3$, we ran several experiments, drawn on a large dataset of match scores that we collected from public sources.%
\footnote{Data were extracted from \url{http://www.lfp.es/LigaBBVA/Liga_BBVA_Resultados.aspx}}
%
Our dataset contains the complete scores of several seasons of the  Spanish {\em Primera Divisi\'{o}n} (\emph{Liga}), the top football league in Spain. 
At the moment of writing, 20 clubs play in the Liga. 
Each club plays every other club twice, once at home and once when visiting the other club.
Points are assigned according to the $3/1/0$ schema: 3 for win, 1 for draw and 0 for loss.
Clubs are ranked by the total number of points they accumulate and the highest-ranked club at the end of the season is crowned champion. 
The dataset consists of 8182 matches from the 1928-29 season until the 2011-12 season. 
Overall, the home club won 3920 times and lost 2043 times, and the number of ties amounted to 2119.
 

For the football domain, a major goal of our experimental tests was to check MORE's predicting accuracy.
For each match in our database involving clubs $\alpha$ (home club) and $\beta$ (visiting club) we separately applied the {\em Naive,} {\em MV} and {\em GMV} strategies to convert the outcomes of a football match into opinions.
We then applied the MORE algorithm and computed the relative strength $r_{\alpha, \beta}$ of $\alpha$ against $\beta$.  
We tried various configurations of the decay parameter $\nu$ in order to study how the tuning of this parameter influences the overall predictive performance of MORE. 
The usual $3/1/0$ scoring system for football rankings (and other games) 
provided us with a baseline to study the predictive accuracy of MORE.

The experimental procedure we followed to compare the predictive accuracy of MORE and $3/1/0$ was as follows.  
We partitioned the dataset containing football matches into 10 intervals, $\mathcal{I}_1, \mathcal{I}_2, \ldots, \mathcal{I}_{10}$, on the basis of the relative strength of opponent clubs.
In detail, for an arbitrary pair of clubs $\alpha$ and $\beta$, the first interval $\mathcal{I}_1$ was formed by the matches such that $0 \leq r_{\alpha, \beta} < 0.1$, the second interval $\mathcal{I}_2$ contained the matches for which $0.1 \leq r_{\alpha, \beta} < 0.2$ and so on until the tenth interval $\mathcal{I}_{10}$ (consisting of the matches in which $0.9 \leq r_{\alpha, \beta} \leq 1$).
Observe that the intervals may have different sizes (because, for instance, the number of matches in $\mathcal{I}_1$ could differ from those in $\mathcal{I}_2$).
Given an interval $\mathcal{I}_k$, we have that the larger $k$, the better the skills of $\alpha$ are and, then, the more likely $\alpha$ should be able to beat $\beta$.

For different strategies and parameter settings, we computed the percentage of times ($F_H(k)$) that MORE accurately predicted the outcome of matches in the $\mathcal{I}_k$ interval that ended with the victory of the home club.  Accordingly, we refer to $F_H(k)$ as the {\em home success frequency}.
In an analogous fashion, we computed the percentage of times ($F_{A}(k)$) that MORE accurately predicted the outcome of matches in the $\mathcal{I}_k$ interval that ended with the victory of the visiting club.  Accordingly, we refer to $F_A(k)$ as the {\em visiting success frequency}.  
We would expected that the higher $r_{\alpha,\beta}$ the higher $F_{H}(k)$. 
In fact, as $r_{\alpha,\beta} \rightarrow 1$ MORE becomes more and more confident on the ability of $\alpha$ of beating $\beta$ and, therefore, we expect that $F_{H}(k)$ is consequently large. 
The situation for $r_{\alpha,\beta}$ is similar: its increase corresponds to a decrease of $r_{\beta,\alpha}$ and, therefore, an increase of $r_{\beta,\alpha}$ should correspond to a decrease in the frequency of (home club) $\alpha$ wins.



In the following, when it does not generate confusion, we shall use the simplified notation $F_{H}$ (resp., $F_{A}$) in place of $F_{H}(k)$ (resp., $F_{A}(k)$) because, for a fixed match $\langle \alpha, \beta \rangle $ we can immediately identify the interval $\mathcal{I}_k$ to which $\langle \alpha, \beta \rangle $ belongs to and, therefore, the $F_{H}(k)$ (resp., $F_{A}(k)$) becomes redundant.

\subsection{Assessing the Quality of Predictions}\label{sub:quality-fmore}
\begin{figure*}[!b]
       \centering
				\parbox{0.5\textwidth}{\includegraphics[width=0.45\textwidth]{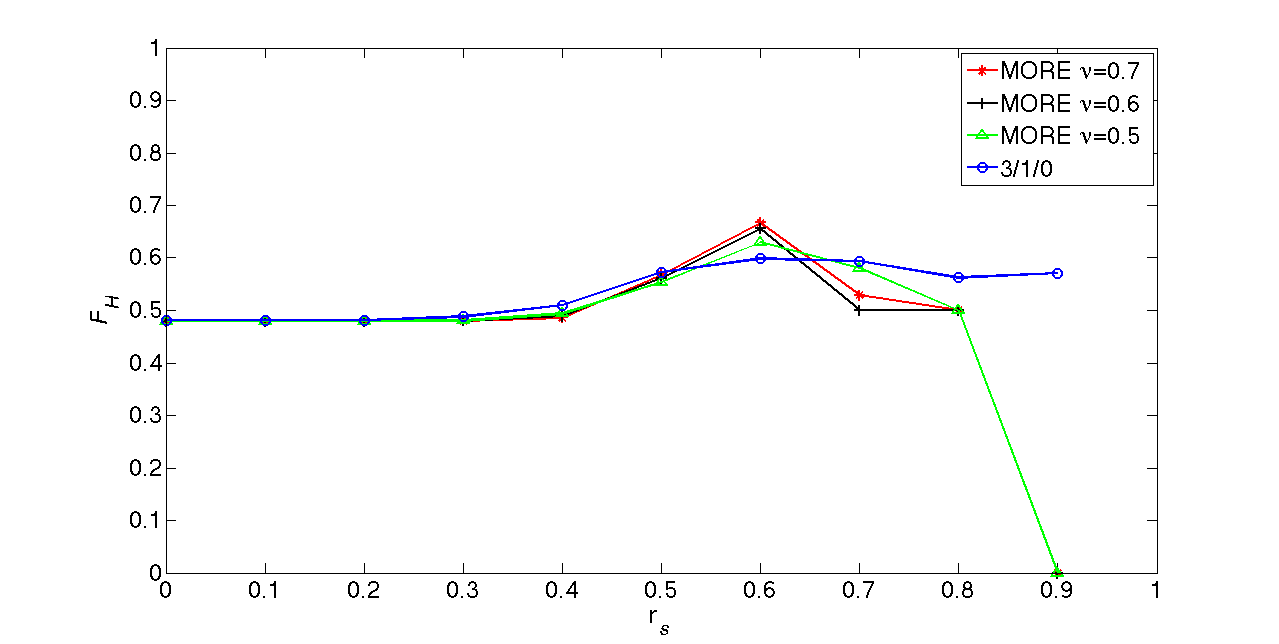}}%
				\parbox{0.5\textwidth}{\includegraphics[width=0.45\textwidth]{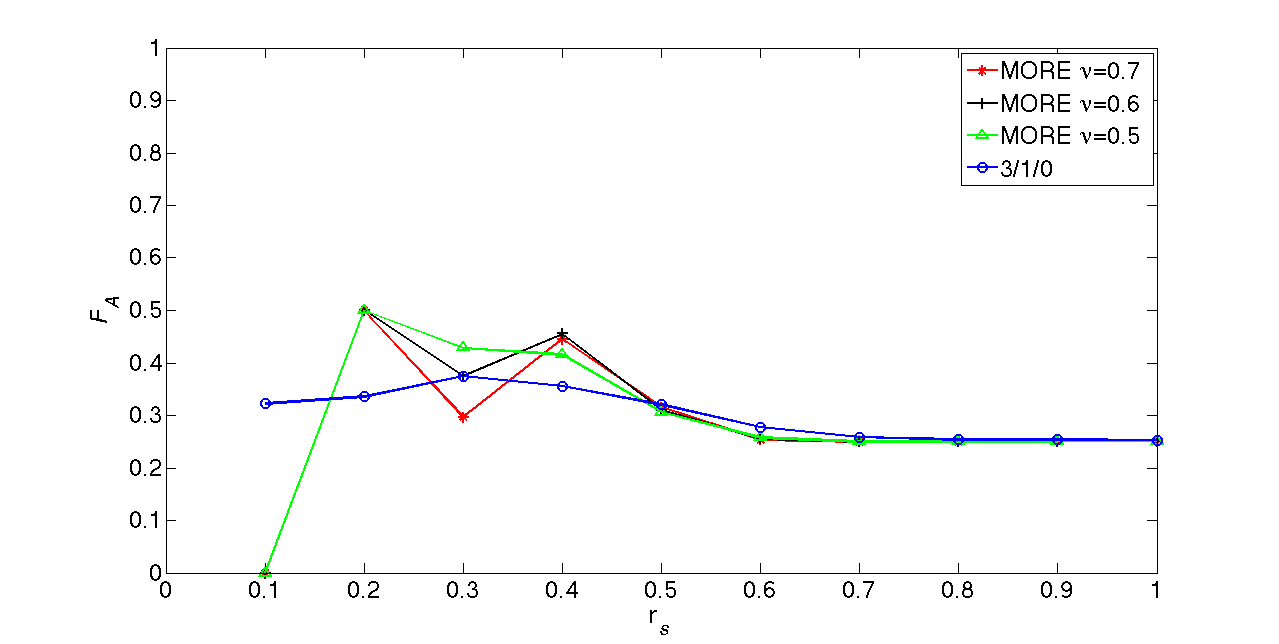}}
        \caption{Naive strategy: success frequencies for $F_H$ and $F_A$ (resp.) over relative strength.}
        \label{fig:Naive-res}
\end{figure*}

\begin{figure*}[!b]
       \centering
				\parbox{0.5\textwidth}{\includegraphics[width=0.45\textwidth]{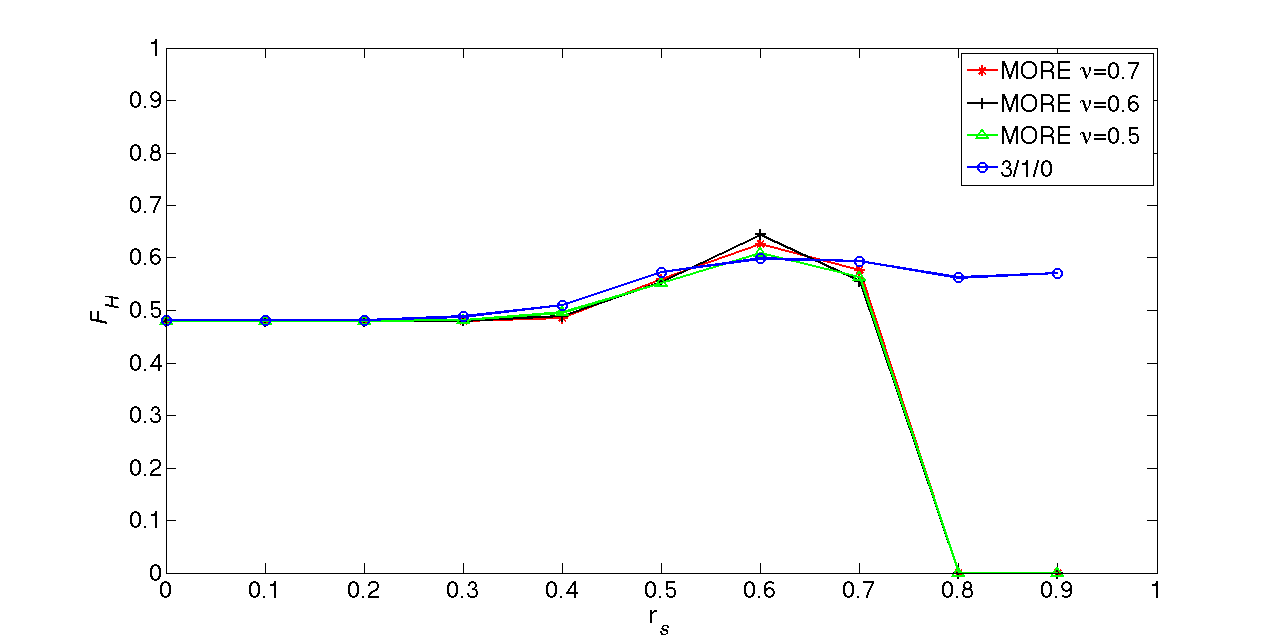}}%
				\parbox{0.5\textwidth}{\includegraphics[width=0.45\textwidth]{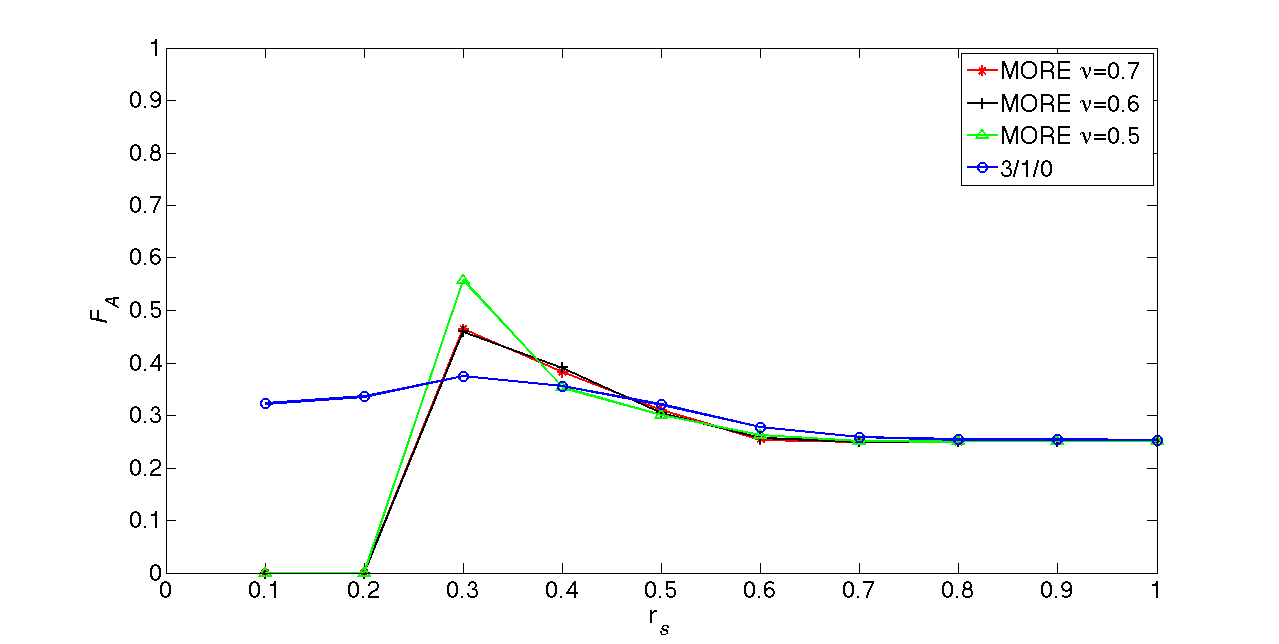}}
        \caption{MV strategy: success frequencies for $F_H$ and $F_A$ (resp.) over relative strength.}
        \label{fig:MV-res}
\end{figure*}

\begin{figure*}[!b]
       \centering
				\parbox{0.5\textwidth}{\includegraphics[width=0.45\textwidth]{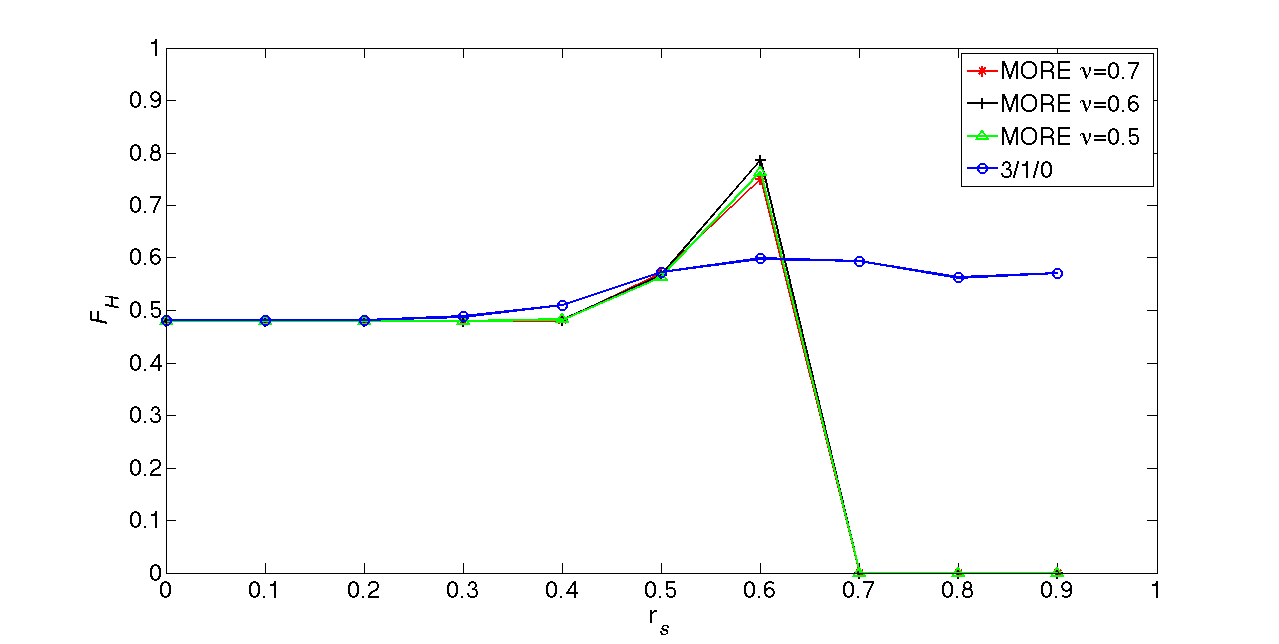}}%
				\parbox{0.5\textwidth}{\includegraphics[width=0.45\textwidth]{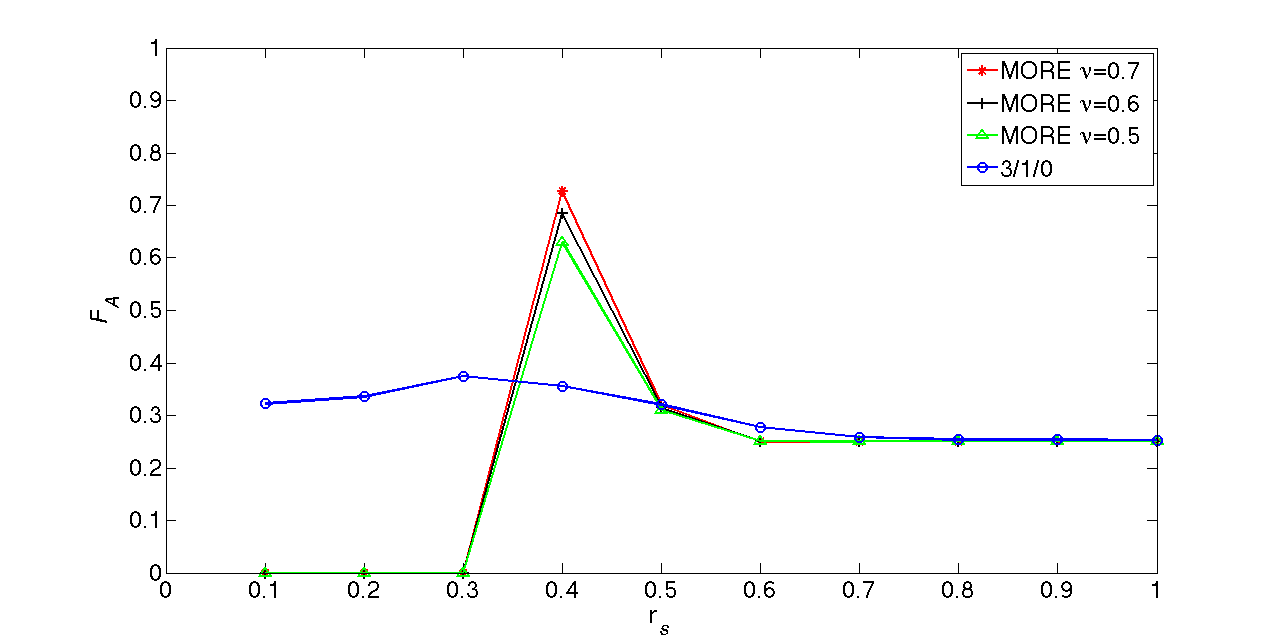}}
        \caption{GMV strategy: success frequencies for $F_H$ and $F_A$ (resp.) over relative strength.}
        \label{fig:GMV-res}
\end{figure*}

The first series of experiments we performed aimed at assessing the accuracy of the predictions with respect to the different strategies.
The results are plotted in Figures~\ref{fig:Naive-res}--\ref{fig:GMV-res} for the Naive, MV, and GMV strategies, respectively. 
In each figure, the plot on the left represents the frequency of successful predictions for the home team winning, and that on the right represents the frequency of successful predictions for the visiting team winning. 



From the analysis of these results we can draw some relevant conclusions.  
The Naive strategy, despite its simplicity and independence from the final outcome of the match, is able to generate accurate predictions. 
In fact, Naive is ofter a better forecaster than ranking-based prediction (i.e., using the 3/1/0 point system).  
For home victories, the maximum value of $F_H$ is around 66\% whereas the 3/1/0 algorithm peaks at around 59\%.
For away victories, the values of $F_A$ range between 25\% and 50\% whereas the $3/1/0$ algorithm has its success frequency flat around 0.3. 

It is also interesting to observe that the decay factor $\nu$ has little impact on the values of both $F_H$ and $F_A$. 
In particular, the peak value of $F_H$ is obtained when $\nu = 0.7$ but the value $\nu = 0.5$ provides more stable results.
In contrast, setting $\nu = 0.5$ is the best option for visiting victories, even if the curves describing the evolution of $F_A$ tend to coincide when the relative strength (depicted as $r_{s}$ in Figures ~\ref{fig:Naive-res}--\ref{fig:GMV-res}) is greater than 0.5.

Let us now consider the {\em MV} strategy, whose results are reported in Figure~\ref{fig:MV-res}.  
This second experiment provides evidence of an increase in the accuracy of MORE, as the highest value of $F_H$ is now equal to 64\% and the highest value of $F_A$ is equal to 46\%.  
This suggests that including the number of goals scored by each team in the process of generating opinions is effective in better computing the strength of each club and, ultimately, in producing more accurate predictions.  
From these figures we can also conclude that for both home and visiting victories, $F_H$ and $F_A$ achieve their peak when $\nu= 0.6$. But the trends of the curves depicted in Figure\ref{fig:MV-res} are quite similar. This implies that information decay has little impact when the {\em MV} strategy is chosen.

Finally, we consider the {\em GMV} strategy.   
Once again, we computed $F_H$ and $F_A$ for different values of $\nu$ and the corresponding results are graphically reported in Figure \ref{fig:GMV-res}.

This last experiment illustrates that the {\em GMV} strategy (with $X = 1$) provides the highest values of $F_H$ and $F_A$.  
The best value of $F_H$ is around 78\% (while $F_H$ associated with the $3/1/0$ algorithm does not exceed 59\%).  
Analogously, in case of visiting victories, the best value of $F_A$ is equal to 68\% (while the $3/1/0$ algorithm is not able to go beyond 37\%).

The value of $\nu$ providing the peak values of $F_H$ and $F_A$ was 0.6 even though the information decay has little impact, as in the case of {\em MV} strategy.


We conclude this section by observing that when $r_{\alpha, \beta}$ is less than 0.3, the value of $F_H$ is around 0.5, independently of the adopted strategy.  
This result is clearly superior to a merely guess-and-check strategy, where choices are chosen uniformly at random and the probability of guessing the correct result is $\frac{1}{3}$ (as there are three possible outcomes: $\alpha$ winning, $\beta$ winning, or neither - having a draw).
\section{Conclusion}\label{sec:conclusion}
This paper proposed a reputation model based on a probabilistic modelling of opinions, a notion of information decay, an understanding that the reputation of an opinion holder provides an insight on how reliable his/her opinions are, as well as an understanding that the more certain an opinion is, the more its weight, or impact. 

An interesting aspect of this model is that it may be used in domains rich with explicit opinions, as well as in domains where explicit opinions are sparse.  In the latter case, implicit opinions are extracted from the behavioural information.  This paper has also proposed an approach for extracting opinions from behavioural information in the sports domain, focusing on football in particular.  

In the literature, several ranking algorithms exist that are also based on the notion of implicit opinions.  
For instance, PageRank~\cite{pagerank} and HITS~\cite{hits} compute the reputation of entities based on the links between these entities.  
Indirectly, their approach assumes that a link describes a positive opinion: one links to the ``good'' entities. 
Both have been applied successfully in the context of web search.  
In~\cite{expertFinderEg1}, ranking algorithms like PageRank and HITS were applied to the social network to find experts in the network based on who is replying to the posts of whom.  
In~\cite{expertFinderEg2}, HITS has been used in a similar manner to help find experts based on who is replying to the emails of whom.  
EigenTrust~\cite{eigenTrust} calculates the reputation of peers in P2P networks by relying on the number of downloads that one peer downloads files from another.  In~\cite{p2pRep}, a personalised version of PageRank that also relies on the download history is used to find trustworthy peers in P2P networks.
Also, CiteRank~\cite{citerank} and SARA~\cite{sara} are algorithms that rank research work by interpreting a citation as a positive opinion about the cited work.  


In comparison, we note that MORE is more generic than existing ranking algorithms, since it has the power to incorporate both explicit and implicit opinions in one system.  Although built upon previous work, MORE also introduces the novel idea of considering the certainty of an opinion as a measure of its weight, or impact, when aggregating the group members' opinions. Finally,  the model is validated by evaluating its performance in predicting the scores of football matches.
%
%
%
We consider the football league scenario particularly interesting because it describes well the opportunities and limitations of the mechanisms by which we would like to evaluate reputation, and thus estimate the true strength of agents in general. 
%
Furthermore, we note that unlike the sophisticated predictive models in use today, (e.g., Goldman Sachs' model that was used for World Cup 2014, and relied on around a dozen statistical/historical parameters),
MORE relies solely on game scores.  
In other words, it requires no tuning of complex parameters, 
and yet its predictions are reasonably accurate.

\section*{Acknowledgements}
This work is supported by the Agreement Technologies project (CONSOLIDER CSD2007-0022, INGENIO 2010) and the PRAISE project (EU FP7 grant number 388770).

\bibliographystyle{splncs03}
\bibliography{references}

\end{document}